\documentclass[preprint,12pt]{elsarticle}

\usepackage{amssymb}
\usepackage{amsmath}

\usepackage{hyperref}

\usepackage{upgreek}

\usepackage{graphics}

\journal{Optics Communications}

\begin{document}

\begin{frontmatter}



\title{Hard X-Ray Zernike-Type Phase-Contrast Imaging with a Two-Block Crystal System}


\author{Levon Haroutunyan} 
\ead{levhar@ysu.am}


\affiliation{organization={Yerevan State University}, 
            city={Yerevan},
            country={Armenia}}
            


\begin{abstract}

A novel scheme for Zernike-type hard X-ray phase-contrast imaging is proposed. The scheme relies on X-ray dynamical diffraction in a two-block crystal system with parallel crystal plates of equal thickness. A phase shifter providing a $\pi/2$ phase shift is placed in the inter-block gap of the crystal system. The method operates in a scanning geometry. The proposed imaging setup is compact and does not require conventional focusing optics. Numerical simulations of phase-contrast image formation are performed.
\end{abstract}




\end{frontmatter}




\section{Introduction}

As early as the 1970s, it was predicted that a two-block crystal system with parallel crystal plates of equal thickness, cut in the symmetric Laue diffraction geometry (LL system), could act as a diffraction lens for hard X-rays \cite{LL1}. A narrow X-ray beam incident on the first plate at the Bragg angle is dynamically diffracted and as a result spreads into a fan-shaped pattern with an opening angle of $2\theta_\text{B}$, where $\theta_\text{B}$ is the Bragg angle. The broad X-ray beam formed at the exit surface of the first plate is, upon diffraction in the second plate, partially focused at the exit surface of the second plate, while the remaining part continues to diverge, forming a high background. Experimental confirmation of this focusing effect was reported in \cite{LL2}. In \cite{LL3}, the transfer of an X-ray image from the entrance surface of the first crystal plate to the exit surface of the second plate was demonstrated experimentally. In \cite{LL4}, numerical simulations show the efficiency of a scanning technique for improving image transfer quality, in particular for suppressing the above-mentioned background. In \cite{LL5}, the LL system was used to enhance the visibility of the phase-contrast image obtained with a three-block Laue interferometer \cite{LL6, LL7}.

In the Zernike phase-contrast method \cite{z9}, the diffracted and undiffracted components of X-ray radiation, formed as the beam passes through a phase object, are spatially separated. This separation makes it possible to introduce a phase shift of $\pi/2$ between the diffracted and undiffracted radiation by means of a phase shifter. The spatial separation of these components is realized in the back focal plane of the objective, which follows from the well-known Fourier-transforming property of lenses \cite{z10}.

Zernike X-ray phase-contrast imaging was first realized in \cite{z11} for soft X-rays and in \cite{z12} for hard X-rays. In both cases, Fresnel zone plates \cite{z13} were used as both condensers and objective lenses. The radiation incident on the sample has the form of a hollow cone, and the phase shifter, placed in the back focal plane of the objective, is implemented in the form of a ring.

In the present work, an alternative Zernike-type scheme for hard X-ray phase-contrast imaging is proposed. It is based on the image-transfer capability of the LL system and does not rely on conventional focusing optics, such as Fresnel zone plates. The spatial separation of the deflected and undeflected beams, generated as the beam passes through the investigated sample, is achieved due to a feature of dynamical X-ray diffraction in crystals, according to which a small angular deviation (of the order of an arcsecond) of the beam incident on a crystal leads to a deflection of the corresponding diffracted beams by angles of the order of several degrees \cite{Authier}.

\section{Device scheme}

\begin{figure}[htbp] 
\centering
\includegraphics{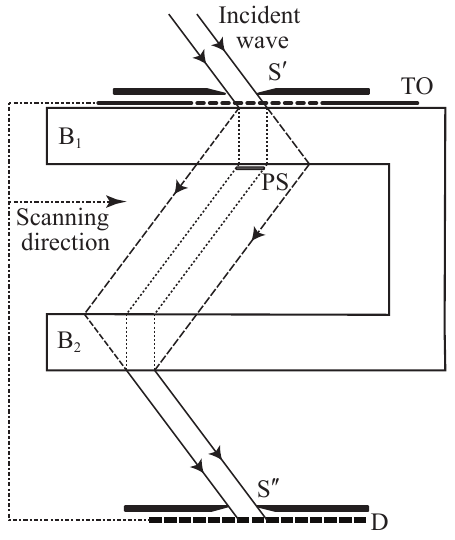}
\caption{Schematic layout of the phase-contrast imaging setup and the corresponding ray paths. The dotted lines indicate the boundaries of the wave packet not diffracted at the entrance slit and the test object, while the dashed lines show the boundaries of the diffracted wave packet. $\text{S}^{\prime}$ and $\text{S}^{\prime \prime}$ are the entrance and exit slits, respectively; PS is the phase shifter; TO is the test object; D is the image detector; $\text{B}_1$ and $\text{B}_2$ are the blocks of the LL system. Note that in a real experimental setup, the slits $\text{S}^{\prime}$ and $\text{S}^{\prime \prime}$, the phase shifter PS, the image detector D, and even the sample TO may be arranged perpendicular to the X-ray beams incident on them, with appropriate scaling of their widths.}
\label{fig:setup}
\end{figure}

The schematic diagram of the proposed device is shown in Fig.~\ref{fig:setup}. The test phase object (TO) is placed between the entrance slit ($\text{S}^{\prime}$) and the entrance surface of the first crystal block of the LL system. A $\pi/2$ phase shifter (PS) is positioned downstream of the exit surface of the first crystal block and is oriented perpendicular to the slit $\text{S}^{\prime}$. The thickness of the phase-shifting plate is determined taking into account the oblique incidence of the radiation, causing the corresponding increase in the optical path length within the plate. The phase-contrast image is recorded in the beam doubly reflected by the crystal blocks, downstream of the exit slit $\text{S}^{\prime \prime}$.

An incident plane wave, aligned along the exact Bragg direction, after passing through the entrance slit $\text{S}^{\prime}$ and the test phase object TO, is incident on the first crystal block. The portion of the beam that is not diffracted by the TO and the slit $\text{S}^{\prime}$ propagates inside the crystal along the normal to the entrance surface and reaches the phase shifter PS, whereas the diffracted portion propagates in an oblique direction with respect to the normal and bypasses the phase shifter. In the second crystal block, these beams converge into foci corresponding to each point of the test object. As a result, a Zernike phase-contrast image of the test object is formed at the detector. In this configuration, a scanning imaging scheme is employed, which, in particular, enables suppression of the diverging radiation by the exit slit $\text{S}^{\prime \prime}$ \cite{LL4}.

Based on geometrical considerations, the widths of the slits $\text{S}^{\prime}$ and $\text{S}^{\prime \prime}$ and that of the phase shifter PS are chosen to be equal. This common width, denoted by $a$, is chosen taking into account the following considerations:

\begin{enumerate}[(a)]
\item\label{it:first} the width of the phase shifter must be sufficiently small so that the bulk of the beam deflected by the test object does not intersect the phase shifter due to its oblique propagation with respect to the normal to the crystal entrance surface;

\item\label{it:second} inhomogeneities of the test phase object with characteristic sizes exceeding the width of the entrance slit cannot be registered by the considered diffraction method, since such objects introduce only a constant phase term into the radiation transmitted through the entrance slit–sample system. Thus, the width of the entrance slit sets an upper limit on the size of detectable inhomogeneities. From this point of view, the entrance slit should be as wide as possible;

\item\label{it:fird} due to specific features of X-ray diffraction in LL systems, the use of a scanning scheme with slit widths of several tens of µm leads to a strong suppression of image background \cite{LL4}.

\end{enumerate}

According to these factors, the calculations yield a width of $a = 32.7 \, \upmu\text{m}$ (see \ref{app:Append}). Here and below, we consider the case of Si(220) reflection of $\text{Mo}K\upalpha$ radiation; the crystal block thickness is $d = 450 \, \upmu\text{m}$.

{\sloppy
The spatial resolution of the considered method is determined by the half-width of the focal peak of the LL system. It is given by $\Lambda \tan \theta_\text{B} \ln 2 / \pi$ \cite{LLZ17} and, under the considered conditions, equals $1.51 \, \upmu\text{m}$. Here, $\Lambda = \lambda \cos \theta_\text{B} / (C |\chi_{h\text{r}}|)$ is the extinction length \cite{Authier}, $\lambda$ is the X-ray wavelength, $\theta_\text{B}$ is the Bragg angle, $C$ is the polarization factor, and $\chi_{h\text{r}}$ is the Fourier component corresponding to the considered reflection of the real part of the crystal polarizability. We also note that, unlike the absorption-contrast imaging scheme \cite{LL4}, in the present case the size of inhomogeneities is also limited from above by the width of the entrance slit.  
\par}

\section{Numerical simulation}

\begin{figure}[htbp] 
\centering
\includegraphics[width=0.9\textwidth]{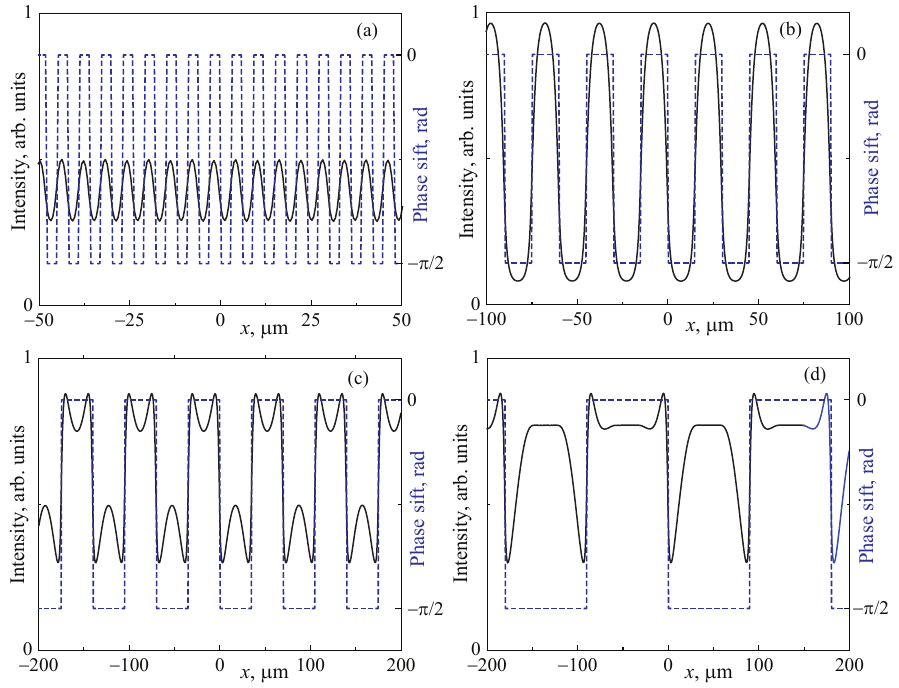}
\caption{Numerical simulation of phase-contrast imaging for a test object in the form of a one-dimensional binary phase grating with a phase jump of $\pi / 2$ and periods of (a) 6, (b) 30, (c) 70, and (d) $180 \, \upmu\text{m}$. Solid lines represent the phase-contrast images (left scale), while dashed lines show the phase-shift distributions of the test object (right scale). The $x$-axis corresponds to the coordinate in the scattering plane, parallel to the surfaces of the crystal plates. The slight image shift in panel (a) is due to the finite pixel resolution of the detector.}
\label{fig:simul}
\end{figure}

{\sloppy
Numerical simulations of Zernike-type phase-contrast imaging implemented with the proposed setup were carried out. A one-dimensional binary phase grating with a phase jump of $\pi / 2$ was used as the test sample. Dynamical diffraction of X-rays in the blocks of the LL system was described by numerical integration of the Takagi equations \cite{LLZ18} using the CSA algorithm \cite{LLZ19}. Both the phase shifter and the test object are assumed to be non-absorbing to highlight the pure phase-contrast effect. The simulation results for different periods of the test phase grating are shown in Fig.~\ref{fig:simul} (solid line). For comparison, the spatial distribution of the phase shift of the test sample is also shown here (dashed line).
\par}

As can be seen from the figure, when the size of a single step (in our case, equal to half of the modulation period) is much smaller than the width of the entrance slit (Figs.~\ref{fig:simul}a and \ref{fig:simul}b), good quality of phase-contrast imaging is observed. In the case of Fig.~\ref{fig:simul}c, where the step size slightly exceeds the width of the entrance slit, the imaging quality deteriorates noticeably. In the case of Fig.~\ref{fig:simul}d, where the step size exceeds the slit width by nearly a factor of three, only regions in the vicinity of the phase-jump points are registered, while the images of individual steps disappear, as noted earlier.

\section{Conclusion}

An alternative Zernike-type scheme for hard X-ray phase-contrast imaging is proposed. The scheme is based on dynamical diffraction of X-rays in an LL system. Although no conventional Fourier-transforming optics are used, the method follows the basic Zernike phase-contrast principle, namely the spatial separation and phase shifting between deflected and undeflected ray components transmitted through the sample. Owing to the geometry of the setup and the high background inherent to image transfer in an LL system, a scanning imaging scheme is employed. In the considered phase-contrast imaging scheme, this limits the maximum size of detectable inhomogeneities by the width of the entrance slit. The setup has a compact layout and does not require conventional focusing optics.

\appendix
\section{Selection of the common width of the entrance and exit slits and the phase shifter}
\label{app:Append}

This common width is selected according to conditions (\ref{it:first})--(\ref{it:fird}).

Considering the upper limit on the characteristic sizes of the detectable inhomogeneities of the test sample, discussed in item (b), we restrict the spatial spectrum of the detectable inhomogeneities of the test object by the inequality $|\omega| \le \omega_0$, where $\omega_0 = 2 \pi / a$, $\omega$ is the spatial frequency of the inhomogeneities, and $a$ is the slit width.

Before reaching the first crystal block, the wave diffracted by the sample inhomogeneities with $|\omega| = \omega_0$ is deflected from the incident-wave direction by an angle

\begin{equation}
\Delta\theta = \lambda / (a \cos \theta_\text{B}),
\label{eq:dTheta}
\end{equation}

\noindent where $\lambda$ denotes the X-ray wavelength and $\theta_\text{B}$ the Bragg angle.

The angular deviation ($\varphi$) of the propagation direction of the corresponding crystal wave from the normal to the entrance surface of the crystal block is defined by \cite{Authier}

\begin{equation}
\label{eq:Phi}
\tan \varphi = \frac{\eta}{\sqrt{\eta^2 + 1}} \tan \theta_\text{B},
\end{equation}

\noindent where

\begin{equation}
\label{eq:eta}
\eta = \frac{\sin{(2\theta_\text{B}})}{C |\chi_{h\text{r}}|} \Delta\theta.
\end{equation}

\noindent Here, the values of $C$ and $\chi_{h\text{r}}$ are given in the main text of the article. In \eqref{eq:Phi}, it is assumed that the incident vacuum wave is aligned along the exact Bragg direction.

The condition (\ref{it:first}), according to which the considered crystal wave packet bypasses the phase shifter, is expressed as

\begin{equation*}
a \le d \tan(\varphi(a))
\end{equation*}

\noindent where the dependence $\varphi(a)$ is defined by Eqs.~\eqref{eq:dTheta}--\eqref{eq:eta}.

Numerical solution of this inequality for the considered experimental parameters yields the condition

\begin{equation*}
a \le 32.7 \, \upmu\text{m} 
\end{equation*}

\noindent Taking into account conditions (\ref{it:second}) and (\ref{it:fird}), the maximum allowable value is chosen for the widths of the slits and the phase shifter, namely

\begin{equation*}
a = 32.7 \, \upmu\text{m}.
\end{equation*}


\end{document}